# Dynamically switching the surface electronic and electrostatic properties of indium tin oxide electrodes with photochromic monolayers

*Qiankun Wang, Valentin Diez-Cabanes, Simone Dell'Elce, Andrea Liscio, Björn Kobin, Hong Li, Jean-Luc Brédas, Stefan Hecht, Vincenzo Palermo, Emil J.W. List-Kratochvil, Jérôme Cornil, Norbert Koch[*] and Giovanni Ligorio[†]*

Q. Wang, Prof. N. Koch

Institut für Physik & IRIS Adlershof, Humboldt-Universität zu Berlin,
Brook-Taylor-Str. 6, 12489 Berlin, Germany

Prof. N. Koch

Helmholtz-Zentrum Berlin für Materialien und Energie GmbH,
Albert-Einstein Str. 15, Berlin 12489, Germany

Prof. E. J.W. List-Kratochvil, Dr. G. Ligorio

Institut für Physik, Institut für Chemie & IRIS Adlershof, Humboldt-Universität zu Berlin, Brook-Taylor-Str. 6, 12489 Berlin, Germany

V. Diez-Cabanes, Prof. J. Cornil

Laboratory for Chemistry of Novel Materials, University of Mons, Place du Parc 20, B-7000 Mons, Belgium

S. Dell'Elce, Dr. A. Liscio, Dr. V. Palermo

CNR-ISOF - Istituto per la Sintesi Organica e la Fotoreattività, Via P. Gobetti, 101 - 40129 Bologna

Dr. A. Liscio

CNR-IMM - Istituto per la microelettronica e microsistemi, Viale del Fosso del Cavaliere, 110 - 00133 Roma

Dr. B. Kobin, Prof. S. Hecht

Institut für Chemie & IRIS Adlershof, Humboldt-Universität zu Berlin, Brook-Taylor-Str. 2, 12489 Berlin, Germany

Dr. H. Li, Prof. J.L. Brédas

Center for Organic Photonics and Electronics and School of Chemistry and Biochemistry, Georgia Institute of Technology, 901 Atlantic Drive N.W., Atlanta, GA 30332-0400, USA

---

[*] nkoch@physik.hu-berlin.de
[†] giovanni.ligorio@hu-berlin.de

**ABSTRACT**

The chemical modification of electrodes with organic materials is a common approach to tune the electronic and electrostatic landscape between interlayers in optoelectronic devices, thus facilitating charge injection at the electrode/semiconductor interfaces and improving their performance. The use of photochromic molecules for the surface modification allows dynamic control of the electronic and electrostatic properties of the electrode and thereby enables additional functionalities in such devices. Here, we show that the electronic properties of a transparent indium tin oxide (ITO) electrode are reversibly and dynamically modified by depositing organic photochromic switches (diarylethenes) in the form of self-assembled monolayers (SAMs). By combining a range of surface characterization and density functional theory calculations, we present a detailed picture of the SAM binding onto ITO, the packing density of molecules, their orientation, as well as the work function modification of the ITO surface due to the SAM deposition. Upon illumination with ultraviolet and green light, we observe a reversible shift of the frontier occupied levels by 0.7 eV, and concomitantly a reversible work function change of ca. 60 meV. Our results prove the viability of dynamic switching of the electronic properties of the electrode with external light stimuli, which could be used to fabricate ITO-based photo-switchable optoelectronic devices.

# 1. INTRODUCTION

The deposition of chemisorbed self-assembled monolayers (SAMs) is a powerful and universal approach to fine tune the desired electronic and electrostatic properties at surfaces and interfaces.[1–3] As a result of SAM formation, molecules are covalently bonded to the surface in an ordered manner and form a homogenous coating with controlled molecular-height thickness. When the deposited molecules possess a permanent dipole moment or chemisorption leads to the change of the surface electrostatic potential, this 2-D array of molecules allows tuning the surface potential,[1,4] and hence the optimization of the interfacial energy level alignment with regard to charge injection as well as extraction.

On the other hand, the incorporation of photochromic or thermochromic molecules (e.g. diarylethenes, azobenzenes, or dihydropyrenes) into the SAM provides a dynamic method to switch the electronic and electrostatic properties through external stimuli such as light or temperature.[5–7] The underlying mechanism is that such molecules can switch reversibly between two isomeric configurations upon illumination at different wavelengths or upon heat absorption. Switching between the two isomers thereby fundamentally changes the electronic properties of the molecules, e.g. their electron affinity,[8] ionization potential,[9] and dipole moment.[10] Thus, with such physical-chemical properties, photochromic SAMs are expected to influence the development of multifunctional devices.[5,11] Among the families of photo-switches, diarylethenes (DAEs) are able to undergo reversible change of the molecular conjugation (i.e. ring opening/closure reaction) upon light illumination. DAEs are extensively employed as photochromic systems due to their advantages of higher fatigue resistance and thermal stability in both isomeric states.[12,13] DAE-based SAMs were employed recently to modify metal electrodes;[6] and proved to be able to optically modulate the current in organic field-effect transistors (OFET) by changing the energy barriers at the metal/DAE interface. We recently demonstrated that DAE molecules functionalized with a phosphonic acid anchoring group can also be successfully employed for the dynamic modification of the surface of a metal oxide such as ZnO.[14] It is therefore expected that such DAE SAMs will have similar effects on indium-tin oxide (ITO), a material relevant to electronic technologies, which has been extensively used as a transparent electrode in optoelectronic devices.

In this contribution, we investigate the hybrid interface resulting from the deposition on ITO of phosphonic acid DAE (PA-DAE) as SAMs. In particular, we demonstrate that the electronic properties of ITO can be dynamically and reversibly tuned by switching PA-DAE between its open and closed forms (the chemical structure of the two isomers are reported in Figure 1). Following our previous deposition methodology, which ensures a high density coverage,[3,14] the PA-DAE SAM was deposited on the transparent electrode, and its modified surface was characterized by scanning force microscopy (SFM) and water contact angle. The chemical bonding of the phosphonic acid anchoring group was assessed by measuring O 1s core level spectra by means of X-ray photoemission spectroscopy (XPS). The orientation of the PA-DAE molecules on the surface was characterized by X-ray absorption spectroscopy. The work function modification and valence electronic properties were studied by ultraviolet photoemission spectroscopy (UPS). Upon illumination of the SAM, we observe a reversible shift of the frontier occupied/unoccupied levels by ca. 0.7 eV and concomitantly a small change in work function by ca. 60 meV. Such light-induced experimental observations (i.e. frontier level shift, work function change) are further substantiated at the density functional theory (DFT) level. Our findings establish the feasibility of dynamical energy level tuning on ITO electrodes, which could be used to fabricate multifunctional optoelectronic devices based on PA-DAE modified ITO electrodes.

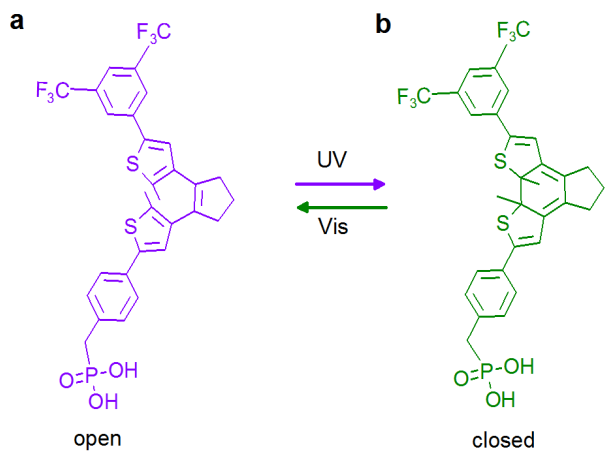

**Figure 1.** Chemical structure of the phosphonic acid diarylethene (PA-DAE) molecule with its (a) open and (b) closed forms, respectively. Upon illumination with UV and visible light, PA-DAE undergoes reversible photoswitching between its open and closed forms.

## 2. METHODS

The synthesis of the PA-DAE molecules is reported in Ref [14]. The deposition of PA-DAE on ITO (roughness < 1 nm, purchased from Thin Film Devices Inc.) is based on a procedure that has been optimized for phosphonic acids and extensively reported previously,[3,14] and is as follows: the SAM was fabricated by immersion of the ITO coated glass into a 1 mM solution of the PA-DAE switch (in the open form) in anhydrous THF for two hours, and subsequently annealed in air for 40 min on a hot plate at 90 °C. Afterwards, sonication in THF for 30 min was carried out to minimized the amount of physisorbed molecules. This preparation procedure was repeated three times in order to achieve a high density of PA-DAE.

Photoemission spectroscopy (PES) and X-ray absorption spectroscopy (XAS) were performed at the SurICat end station of beamline PM4 at BESSY II. For PES, photoelectrons were collected with a hemispherical electron energy analyser in normal emission geometry. The P 2p, C 1s, and O 1s core electron regions were measured with photon energies of 265, 390, and 640 eV, respectively. In this way, the kinetic energy of photoelectrons was kept at ca. 100 eV in order to probe similar sample depth and maximize surface sensitivity. The binding energy of core electrons was referenced to the Au $4f_{7/2}$ level (84.0 eV). The valence electron and core electron regions were characterized with He I ($h\nu$ = 21.2 eV, with Al-foil filter) and Al K$_\alpha$ ($h\nu$ = 1486.6 eV) lines, respectively, in a custom-made system. The secondary electron cutoff (SECO) spectra were recorded with samples biased at −10 V. The Fermi level is referred to as the zero binding energy in all UPS and XPS spectra. Spectra were fitted with both symmetric Voigt and asymmetric Doniach-Sunjic (e.g., for ITO O 1s) profiles after subtracting a Shireley background. XAS experiments were carried out in the total electron yield (TEY) mode. The incident beam was horizontally polarized and the degree of linear polarization was determined to be 0.9. Sample currents were measured using a Keithley 6514 electrometer via a double shielded BNC cable. The sample surface morphology was measured using a scanning force microscope (SFM) equipped with a Bruker Dimension FastScan system in tapping mode. This instrument was also used for Kelvin probe force microscopy (KPFM) investigations.

In order to trigger the switching process, the samples were irradiated *in situ* during UPS measurements through a viewport of the analysis chamber with green (λ centered at 565 nm) or UV (λ centered at 365 nm) light, which were provided by high power LEDs (Thorlabs) mounted with an adjustable collimator. The measured maximum photon intensities were ca. 200 mWcm$^{-2}$ for green light, and 100 mWcm$^{-2}$ for UV light, respectively.

In the case of the DFT calculations, the building of the ITO surface within a slab approach has been reported previously in literature.[15–17] The unit cell consists of three (In-O)/(Sn-O) layers with an In/Sn ratio of 0.14. All oxygen atoms belonging to the top layer of the ITO slab are saturated with hydrogen atoms to model a realistic OH-terminated surface, while the bottom layer is not passivated. The surface unit-cell dimensions are: $a$ = 24.79 Å and $b$ = 14.32 Å; a vacuum region of 32 Å is set in the normal direction to the slabs, with a dipole correction layer introduced into middle (see Figure S1).[18] This vacuum region is large enough to ensure the convergence of the dipole moment in the normal direction (see Figure S2). Four PA-DAE molecules ($n$ = 4) in the open or closed forms were attached to the ITO surface. Note that the adsorption of the PA-DAE molecules is not accompanied by any loss of atoms from the ITO surface; however, the H atoms belonging to the PA group are released upon deposition of the SAM. The O atoms of the PA groups are connected to both OH groups and In/Sn atoms from the ITO surface (see Figure S1). The resulting surface area is A = 88.75 Å$^2$ per molecule. This value is higher than the experimental one obtained by XPS, which is A = 50 ± 11.54 Å$^2$ per molecule. This difference has no major implications in view of a recent work of Li et al. on trifluorophenyl-PA SAMS on ITO showing that, in the high coverage regime ($n$ = 4) of the same unit cell, the value of Δϕ has already converged within 0.1 eV compared to the value obtained with the experimental coverage.[15] The unit-cell optimizations were performed at the DFT level with the PBE exchange-correlation functional[19] within the Gradient Generalized Approximation (GGA) using the plane-wave augmented (PAW) method,[20] as implemented in the Vienna Ab-initio Simulations Package (VASP) code. The plane wave cutoff was set at 300 eV and a value of $10^{-6}$ eV was chosen for the total energy convergence; a $k$-sampling of 2×2×1 and the tetrahedron integration method were used in the Brillouin zone. The geometry optimizations were carried out following a damped molecular dynamics method with a force cutoff for the convergence equal to 0.04 eV Å$^{-1}$. In order to decrease the computational time, the coordinates of the bottom two (In-O)/(Sn-O) layers were frozen during the optimizations.

## 3. RESULTS AND DISCUSSION

### 3.1 Surface morphology

The surface morphologies of the bare ITO and PA-DAE modified ITO were characterized by SFM (see Figure 2). For bare ITO, the height image (Figure 2a) of the surface shows small grains (average size ca. 40 nm in diameter) with a roughness (root mean square) of 0.7 ± 0.1 nm. Upon SAM deposition, the surface roughness does not change significantly (0.4 ± 0.1 nm); the ITO grains are preserved and no island formation or multilayers are observable in the height image (see Figure 2b). Combining the height and the (homogenous) phase images (see Figure 2c and 2d) allows us to conclude that the surface modification results in a uniform coverage. In order to characterize the SAM-modified ITO, we investigated the wettability change induced by PA-DAE by means of a contact angle measurement before and after SAM deposition (water was used as test liquid). The contact angle of water increases significantly from 33.1° ± 1.0° on clean ITO to 97.1° ± 1.0°, as expected from the induced hydrophobic properties of the fluorine head groups attached to DAE. The results are consistent with the observation for PA-DAE SAM on ZnO (where the water contact angle increases from 46° to 101°).[14] Such a wettability change also points towards a densely packed coverage on the ITO surface.[3,21]

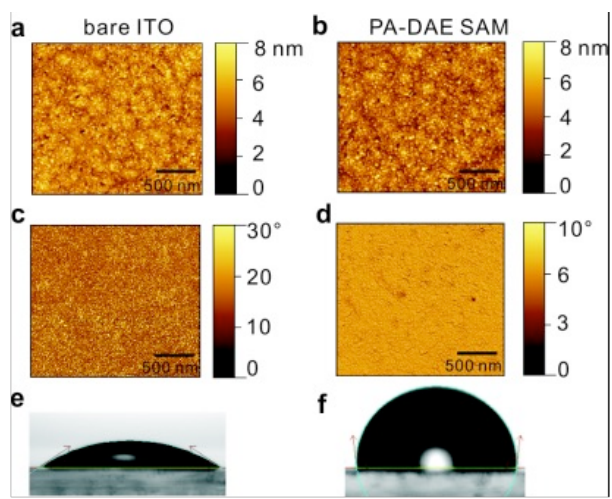

**Figure 2.** SFM height and phase images for bare ITO (a, c) and ITO with PA-DAE SAM (b, d) surfaces. Water contact angles of corresponding surfaces are shown in (e) for bare ITO and (f) for SAM, respectively.

**3.2 Chemical bonding**

To further verify the quality of the SAM films and to confirm chemisorption of the anchoring group of the switching molecules onto ITO, we performed XPS measurements with synchrotron radiation. This enables us to tune the photon energy for photon excitation to ensure a minimal escape depth of the measured electrons and, thus achieve the highest surface sensitivity possible with this technique. The beam damage on the organic molecules during measurement can be neglected by analyzing the C 1s spectra (see Figure S3 and Table S1 in supporting information). The P 2p core level peak was first measured to ensure the presence of phosphonic acid (see Figure 3a). On the bare ITO surface (bottommost spectra), no signal can be detected; however, after surface modification (topmost spectra), the P 2p peak is clearly observed, and two components are resolved from the fitting of the spectra. The component at 133.2 eV in binding energy is attributed to the chemisorbed PA-DAE monolayer.[3] The additional component at higher binding energy (135.0 eV) is attributed to the presence of a small amount of physisorbed molecules (8.0% with respect to the chemisorbed molecules). The packing density of PA-DAE molecules is analyzed by comparing the signal intensity of F 1s with that of In 3d (see Figure S4 and Equation S1 in supporting information), thus yielding ca. 2.0 ± 0.7 molecules nm$^{-2}$. The result is comparable to previous observations for PA-DAE on ZnO.[14]

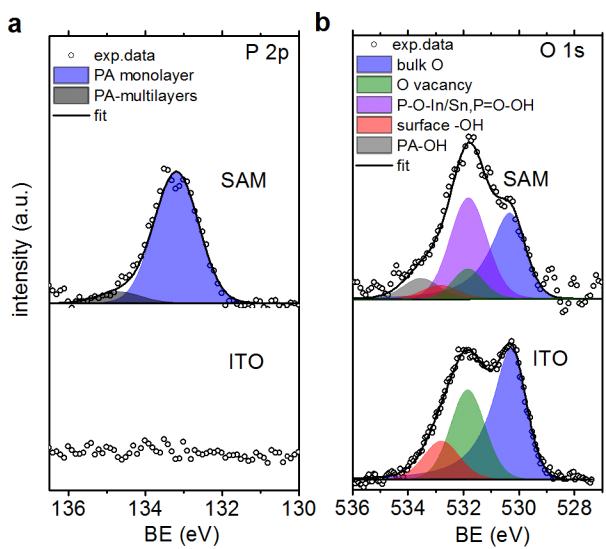

**Figure 3.** P 2p (a) and O 1s (b) core level spectra for bare ITO (bottom) and ITO with PA-DAE SAM (top). The decomposition of the P 2p spectra shows that PA-DAE molecules

primarily form a monolayer; and the decomposition of the O 1s spectra reveals the chemical nature of the phosphonic acid bonding to ITO.

O 1s core level spectra were measured to investigate the nature of the chemical bonding between the DAE phosphonic group and ITO. Figure 3b displays the O 1s spectra before (bottommost) and after (topmost) PA-DAE modification. For bare ITO, three spectral components are resolved. Following earlier DFT calculations,[22] the two components at lower binding energy are attributed to (*i*) ITO bulk oxygen (at 530.3 eV, the peak is asymmetric due to energy loss), and (*ii*) surface oxygen vacancy (at 531.8 eV). The third resolved component is located at 532.8 eV and is attributed to the adsorbed surface hydroxyl groups.[23] After SAM deposition, the O 1s intensity from bulk ITO is markedly attenuated, and the feature at ca. 531.7 eV (1.4 eV higher than the binding energy of bulk oxygen) in the spectral signal increases in intensity. The fitting of the spectrum suggests that an additional component has to be considered in order to justify the intensity modification in this region; this additional peak arises from the chemisorption of the phosphonic anchoring group. In our previous studies, it was established that the chemisorption of PA-DAE on ZnO surfaces occurs in both the bidentate and tridentate binding modes.[14] We thus tentatively attribute the PA peak to a coexistence of bidentate and tridentate coordination on ITO.

To support the assignment of the new O 1s peak and to elucidate the binding mechanism between PA-DAE and ITO, DFT calculations were performed to estimate the O 1s core level shift. Based on literature,[15,24,25] the coexistence of both bidentate and tridentate binding modes in DFT calculations has been considered with a 1:1 ratio. Here, after optimization of the unit cell (see Figure 4a and b) sites 1 and 3 are found to display a bidentate binding with two PA O atoms bonded to Sn/In atoms and another O atom in electrostatic interaction with an ITO surface hydroxyl group; sites 2 and 4 exhibit tridentate binding with the three O atoms of PA bonded to surface Sn and In atoms. The core level shift of the O 1s belonging to the phosphonic acid groups were calculated in the final-state approximation.[26,27] The calculated core level shift of PA with respect to the binding energy of bulk oxygen is shown in Figure 4c. Among the four adsorption sites, we find that the core level binding energies are not highly affected whether the O atom is electrostatically connected to the hydroxyl groups in the bidentate mode or bonded to ITO Sn/In atoms; the majority of the values lie between 1.2 and 1.3 eV for both bidentate and tridentate modes. These values are in quantitative agreement with the XPS data (1.4 eV) discussed in Figure 3b, which points to the fact that bidentate and tridentate modes most likely coexist and are not energetically resolved. Therefore, using the bulk oxygen binding energy from the bare ITO as a reference (530.3 eV from XPS), one peak at ca. 531.7 eV should be added in the deconvolution made in Figure 3 to account for the coexistence of bidentate and tridentate binding. In addition, no clear changes in core level shifts are observed in the DFT calculations when going from the open to closed forms (see Table S2), which implies that photo-switching of the SAM does not impact the PA bonding strength.

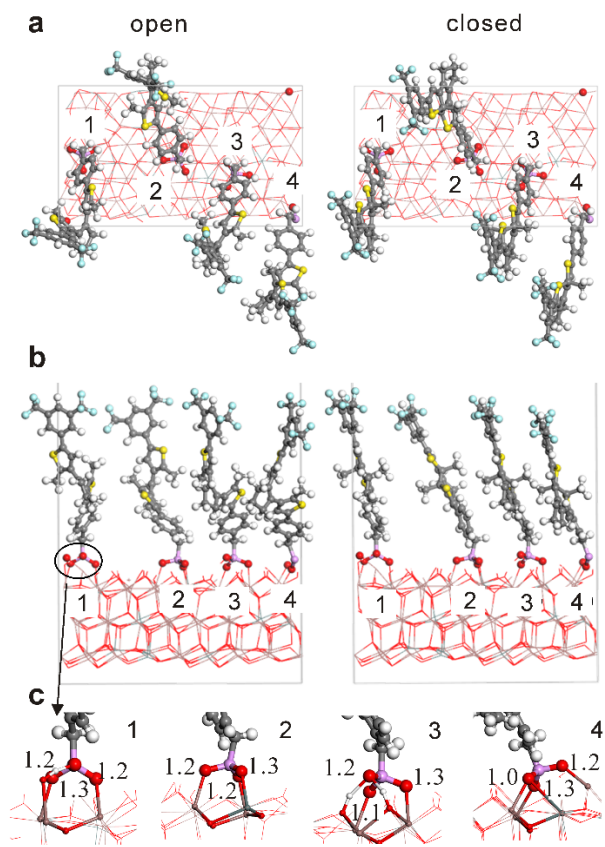

**Figure 4**. Top (a) and lateral (b) views of the ITO unit cells with PA-DAE SAM in the open- (left) and closed- (right) isomeric forms. The figure illustrates the two bidentate (sites 1 and 3) and tridentate (sites 2 and 4) binding modes as well as the zoom on the phosphonic acid binding to ITO for the four different adsorption sites (c); the values (in eV) reported in (c) correspond to the DFT calculated O 1s binding energy shifts with respect to the oxygen atoms in bulk ITO.

### 3.3 Orientation of the SAM

Angle dependent XAS was performed to measure the orientation of the PA-DAE molecules once the SAM is formed on ITO. As shown in Figure 5a, the spectra were collected from the C K-shell electrons excited to unoccupied molecular orbitals. The most intense transition located at 285.0 eV exhibits a strong angular dependence and is assigned to C 1s to π* orbital transitions. Since the C 1s initial state of carbon is spherically symmetric and bears no angular momentum, the angular dependencies of absorption intensity will be directly dependent on the π* orbital orientation in the final state. For the open form, the lowest unoccupied π* molecular orbitals are perpendicular to the top phenyl plane (shown in Figure 5c). To quantify the average orientation of PA-DAE in the SAM, the intensity of π* ($I_{\pi*}$) was measured at five different angles between the incoming polarized light and the sample normal. The molecular tilt angle $\alpha$ (with respect to the ITO surface, see inset of Figure 5b) is calculated as best fit according to equation S2 (for details, see the supporting information). The average tilt angle is thus calculated to be $\alpha = 60°$. Since the molecular orientation in the monolayer is strongly dependent on the molecule-substrate and molecule-molecule interactions as well as on the molecular packing details,[17,25] such results are reasonable when compared to other fluorinated alkyl- and benzylphosphonic acids on ITO.[16,17,25] The measured tilt angle is further supported by the DFT results, which indicate that for the geometrically relaxed monolayer on ITO, PA-DAE displays an average angle of ca. 58° with respect to the surface in both open and closed forms (see Table S3 and Figure S5 for tilt angle details).

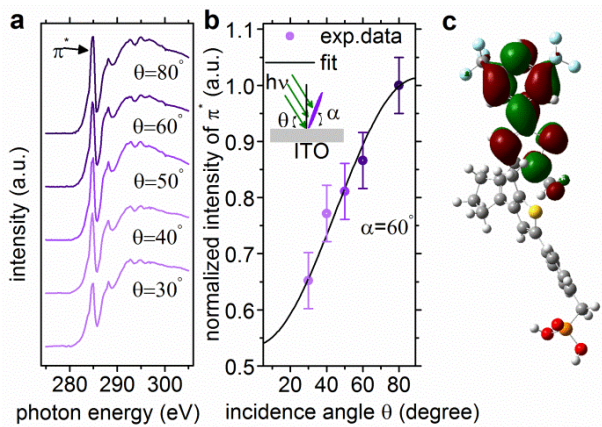

**Figure 5.** Angle dependent C K-edge X-ray absorption spectroscopy (a) of the PA-DAE SAM on ITO. Plots of the $\pi^*$-orbital intensities as a function of the photon incidence angle $\theta$ (b). The solid curve corresponds to the best fit of the intensity evolution indicating with a tilt angle of 60° for PA-DAE. Molecular orbitals for the excited state corresponding to the first $\pi$-resonance of carbon atoms (c).

**3.4 Electronic properties of the SAM**

UPS measurements were performed to investigate the changes induced by PA-DAE on the valence electronic properties of ITO. Figure 6a shows the UPS spectra of the bare ITO and with PA-DAE modification; the latter sample was measured immediately after SAM deposition and thus PA-DAE was initially in the open form. The work function ($\phi$) of the bare ITO was determined by measuring the secondary electron cutoff (SECO) region and amounts to 4.6 eV (see leftmost panel in Figure 6a). The modification with the PA-DAE SAM increases $\phi$ by 0.7 eV up to 5.3 eV. The valence region (central panel) also displays a modification in the density of states upon SAM deposition. The UPS spectra exhibit a structure assigned to the HOMO of open PA-DAE (PA-DAE-o), with the onset measured at 1.1 eV binding energy. Thus, the ionization potential (calculated as the sum of SECO and HOMO) is 6.4 eV, in good agreement with previous results on ZnO.[14] The SAM was subsequently illuminated in situ (for 60 seconds) with UV light to induce a photo-switching from the open to the closed form (PA-DAE-c). The induced PA-DAE-c isomer displays an increased molecular conjugation (see Figure 1), which translates into the appearance of a feature at lower binding energy than the HOMO of molecules in the open configuration. Thus, this feature is the PA-DAE-c HOMO with an onset at 0.4 eV. Subsequent illumination of the SAM with green light for 120 s reversibly switches PA-DAE to its open form, as evidenced by the disappearance of the PA-DAE-c HOMO and reappearance of the PA-DAE-o valence features.

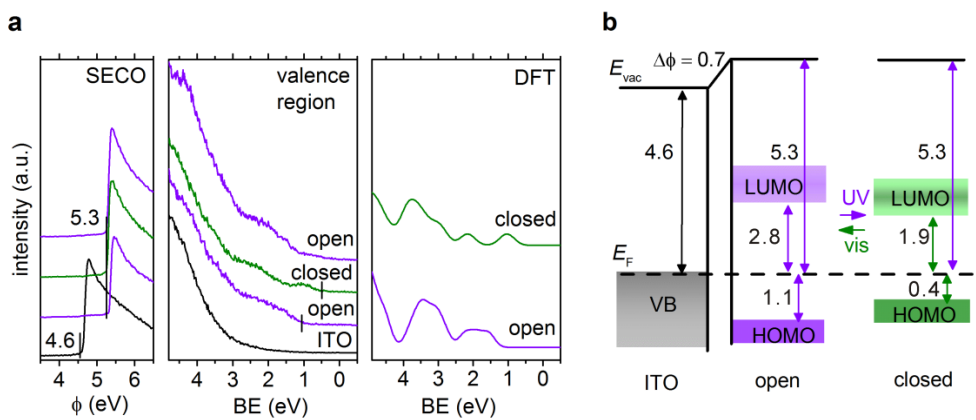

**Figure 6.** Evolution of work function (a, left) and valence region (a, center) of ITO upon deposition of PA-DAE SAM. The figures report the induced changes upon illumination with UV and visible light. DFT calculated density of states (DOS) for PA-DAE for the open and closed isomers are also plotted for comparison (a, right). Schematic energy level diagram at the ITO/PA-DAE interface (b). The energy positions of vacuum level ($E_{vac}$) and HOMO level are derived from UPS, while the LUMO level of PA-DAE is calculated using the transport gap of related DAE derivatives.[8]

To assess the valence electronic features, the density of states of isolated PA-DAE were computed for both open and closed forms and broadened with Gaussian functions (the eigenvalues are convoluted by a Gaussian function with a full width at half maximum of 0.4 eV to be consistent with the UPS resolution). The calculated density of states is in excellent agreement with the UPS data (see rightmost panel in Figure 6a) and confirms the experimental observation of HOMO energy shift upon photo-switching between the isomers.

Interestingly, no clear work function change is observed during the photo-switching process (see SECO). To better assess the work function modification upon switching, the surface potential was mapped with Kelvin probe force microscopy (KPFM). A reversible switching of the surface potential is observed upon UV and green light illumination cycles (Figure S6 and S7). In particular, the PA-DAE-c surface potential turns out to be ca. 60 ± 5 meV higher than the PA-DAE-o surface potential. Similar switching-induced potential changes have been observed in other DAE SAMs on metal by KPFM.[6] We note that the reason why UPS does not resolve such a small potential change can presumably be attributed to: (*i*) possible charging effect during UPS measurements; and (*ii*) UV induced photo-degradation;[28] the latter would give rise to a relatively poor switching yield, the measured work function then likely arising from a superposition of open and closed forms.

The KPFM potential change can be directly attributed to the change in PA-DAE dipole moment ($\mu$) upon switching. It is useful at this stage to revisit the packing density ($n$) of the SAM molecules, which can be estimated according to the Helmholtz equation:[29]

$$(\Delta\phi)_{o-c} = \frac{en(\Delta\mu)_{o-c}\sin(\alpha)}{\varepsilon_r\varepsilon_0} \tag{1}$$

where $(\Delta\phi)_{o-c}$ is the KPFM potential change upon switching, i.e. 60 meV; $e$ is the elementary charge; $\alpha$ is the molecular tilt angle as derived from XAS; $(\Delta\mu)_{o-c}$ is the dipole moment difference between the open and closed forms for the isolated molecules. According to our previous calculations on isolated molecules,[14] $(\Delta\mu)_{o-c}$ is 0.2 D; however, it should be noted that a high molecular coverage (typically for complete or quasi complete monolayers) not only increases the density of dipoles, but also enhances the depolarization effect of surrounding dipoles.[30] This can be accounted for by a relative dielectric constant $\varepsilon_r$ of ca. 1.5.[3,15] The packing density $n$ thus results to be 1.5 ± 0.3 molecules nm$^{-2}$ as calculated with Equation 1, in good agreement with the value of 2.0 ± 0.7 molecules nm$^{-2}$ estimated from XPS, as discussed above.

We summarize in Figure 6b the energy level alignment for ITO functionalized with the PA-DAE SAM. The values of the relative energy separation between Fermi level ($E_F$), HOMO level and vacuum level are derived directly from UPS, while the LUMO energy is obtained by referring to the transport gap of DAE derivatives measured by inverse photoemission spectroscopy.[8] Upon SAM deposition the work function increases with a consequent upshifting of the vacuum level; this is attributed to the formation of interface dipoles, as discussed below. Similar to the observations made on ZnO, PA-DAE can be used to optically control and modulate the hole injection barrier, since the energy difference between HOMO and $E_F$ is reversibly modulated upon illumination with light (i.e. 1.1 eV and 0.4 eV for the open and closed forms, respectively). In parallel, the LUMO to $E_F$ energy difference decreases from 2.8 eV to 1.9 eV when going from open to closed forms. Thus, the HOMO and LUMO energy differences between open and closed forms are $\Delta E_{o-c,HOMO}$ = 0.7 eV and $\Delta E_{o-c,LUMO}$ = 0.9 eV, respectively.

In order to further understand the relative energy level alignment between the two isomers, we have also calculated at the DFT level the frontier level evolution of PA-DAE going from the isolated molecule to the packed monolayer. In the isolated molecules, the HOMO [LUMO] of PA-DAE-c is located at higher [lower] energy than that of PA-DAE-o due to the increase in conjugation and the resulting bandgap reduction in the closed form (see Table S5); the calculated $\Delta E_{o-c,HOMO}$ and $\Delta E_{o-c,LUMO}$ are both 0.7 eV. When increasing the calculated coverage up to a full monolayer, all frontier molecular orbital energies exhibit an upward shift by ca. 0.3 eV (see Figure S8) due to the intermolecular interactions; this shift saturates at a degree of coverage $n = 3$ (i.e., 3 molecules per unit cell). However, the relative energy level alignment remains preserved, i.e. $\Delta E_{o-c,HOMO}$ = 0.5 eV and $\Delta E_{o-c,LUMO}$ = 0.9 eV. After deposition on the ITO surface, the interaction of the SAM with the ITO surface does not significantly affect this relative alignment, with calculated values of $E_{o-c,HOMO}$ = 0.8 eV and $\Delta E_{o-c,LUMO}$ = 0.6 eV (see Figure S9).

**3.5 Theoretical understanding of the surface potential change**

To shed light on the surface potential modification at the ITO/PA-DAE interface, we performed theoretical modeling at the DFT level. The isolated PA-DAE molecules were first geometrically optimized with the PBE functional[19] and four PA-DAE molecules were subsequently grafted on the ITO surface (in both tridentate and bidentate binding modes); the resulting unit cell was then optimized, as described in the Methods section. The work function modification, defined as the work function difference ($\Delta\phi$) between the SAM modified ITO ($\phi_{ITO/SAM}$) and bare ITO ($\phi_{ITO}$), can be decomposed into three contributions:[15,31,32]

$$\Delta\phi = \phi_{ITO/SAM} - \phi_{ITO} = V_{BD} + V_{\mu SAM} + V_{Relax\text{-}ITO} \qquad (2)$$

where $V_{BD}$ represents the creation of a bond dipole at the ITO/PA-DAE interface due to charge reorganization upon SAM formation; $V_{\mu SAM}$ is linked to the intrinsic dipole of PA-DAE in the monolayer; and $V_{Relax\text{-}ITO}$ is the work function change of ITO due to surface geometry relaxation upon SAM grafting. In order to quantify the first term $V_{BD}$ in eq. 2, Poisson's equation $\nabla^2 V(z) = -\Delta\rho(z)/\varepsilon_0$ is used and requires the evaluation of the charge density difference $\Delta\rho(z)$. The latter is calculated in the so-called "molecular scenario" as:[33,34]

$$\Delta\rho(z) = \rho_{ITO/SAM}(z) - \rho_{ITO}(z) - \rho_{PA\text{-}DAE\text{-}H}(z) + \rho_H(z) \qquad (3)$$

where $\rho_{ITO/SAM}$ denotes the plane-averaged charge density of the full ITO/SAM system; $\rho_{ITO}$ the charge density of the relaxed ITO surface; $\rho_{PA\text{-}DAE\text{-}H}$ the charge density of the PA-DAE layer with H atoms added to recover the neutral PA-DAE molecule; and $\rho_H$ the charge density of the isolated H atoms in the same location as in the neutral PA-DAE SAM. The calculations were independently performed with both open and closed isomer configurations. Figure 7 displays $\Delta\rho(z)$ and its cumulated charge reorganization ( $\Delta Q(z) = \int \Delta\rho(z)dz$ ) upon SAM formation along the $z$ axis (taken as the direction normal to the surface).

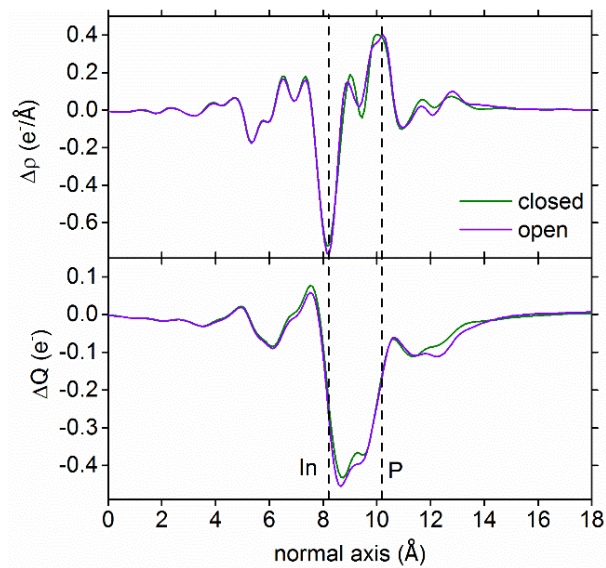

**Figure 7.** Plane-averaged charge density difference (top) and charge transfer (bottom) at the ITO/PA-DAE interface for open- and closed- forms. The vertical dashed lines denote the averaged position of the top layer In and P atoms.

The calculations show that $\Delta\rho$ exhibits a similar distribution for both open and closed forms and that the amount of charge transferred ($\Delta Q$) from ITO to PA-DAE has the same magnitude. A Bader population analysis[35] indicates that the amount of charge transfer is 1.36 |e| per PA-DAE molecule (see Table S4). The transferred charge is mostly localized within the PA anchoring group (-1.26 |e|), while the charges in the DAE backbone are only slightly modified (-0.1 |e|). The estimation of the molecular dipole contribution of the SAM to $\Delta\phi$ ($V_{\mu SAM}$) is made from the isolated monolayer of PA-DAE molecules (i.e. non-bonded to the ITO surface) while keeping the same molecular geometry as in the SAM bonded to ITO (hydrogen atoms were added to keep the system neutral). The plane-averaged electrostatic potential was computed across the SAM layer. The third term in Equation 2 is related to the nuclear relaxation effect in ITO ($V_{\text{Relax-ITO}}$) upon SAM formation; it is evaluated as the difference $V_{\text{Relax-ITO}} = \phi'_{\text{ITO}} - \phi_{\text{ITO}}$ between the work function of the pristine ITO surface ($\phi_{\text{ITO}}$) and work function of the ITO surface without any SAM but in the geometry it adopts upon SAM grafting ($\phi'_{\text{ITO}}$). The work function change and its three components are summarized in Table 1. For both open and closed SAMs, the values of $V_{\text{BD}}$ are compensated by $V_{\text{Relax-ITO}}$ ($V_{\text{BD}} + V_{\text{Relax-ITO}} \approx 0.2$ eV), thus making the magnitude of $\Delta\phi$ primarily driven by the molecular dipole component $V_{\mu\text{SAM}}$. The DFT calculated $\Delta\phi$ for the closed form is ca. 200 meV higher than that of the open form, which is consistent with the KPFM results (60 meV). This $\Delta\phi$ difference between the open and closed forms is mainly ascribed to a higher value of the molecular contribution ($V_{\mu\text{SAM}}$) in the closed form. A further theoretical analysis of the PA-DAE dipole magnitude as a function of coverage (Figure S10 and Table S6) illustrates that the difference in $V_{\mu\text{SAM}}$ between the open and closed forms is slightly reduced when increasing the degree of coverage, due to the different extent of depolarization effects between the two forms.[30] Although the relative changes in $\Delta\phi$ is well reproduced by the calculations, it has to be emphasized that the absolute value of the theoretical $\Delta\phi$ is twice as large as the experimental value (0.7 eV). We suggest that this discrepancy is likely related to a Fermi level pinning effect of the molecular levels, which creates another interfacial dipole contribution. This process appears not to be well depicted by the DFT calculations and two reasons can be put forward for this: (i) the relative energy level alignment of the isolated components (ITO versus PA-DAE molecules) is not adequately described; and/or (ii) much larger unit cells are needed to describe a full charge transfer between the molecule and the electrode to properly describe the Fermi level pinning. Fully addressing this issue is, however, beyond the scope of the present work.

|        | $V_{BD}$ | $V_{\mu SAM}$ | $V_{Relax-ITO}$ | $\Delta\phi = \sum_i V_i$ | $\phi_{ITO/SAM}$ | $\phi_{ITO}$ | $\Delta\phi = \phi_{ITO/SAM} - \phi_{ITO}$ |
|--------|----------|---------------|-----------------|---------------------------|------------------|--------------|--------------------------------------------|
| open   | 0.61     | 1.29          | -0.42           | 1.48                      | 4.68             | 3.20         | 1.48                                       |
| closed | 0.55     | 1.57          | -0.41           | 1.71                      | 4.92             | 3.20         | 1.72                                       |

**Table 1.** Decomposition of the ITO work function modification ($\Delta\phi$) into its three components according to DFT. $\Delta\phi$ is calculated according to $\Delta\phi = V_{BD} + V_{\mu SAM} + V_{Relax-ITO} = \phi_{ITO/SAM} - \phi_{ITO}$ (see Eq. 2), all quantities are defined in the text and are given in eV.

## 4. CONCLUSIONS

In summary, we have demonstrated the possibility of controlling and optically modulating the electrostatic potential landscape (and thus, the electronic energy level alignment) at interfaces with ITO, one of the most widely utilized transparent conductive oxides. For this purpose, the bare surface of ITO was functionalized with a SAM of photochromic diarylethene molecules (functionalized with phosphonic acid as anchoring group). The PA-DAE SAM was comprehensively investigated by contact angle, SFM, XPS, XAS, and UPS measurements. The combination of these characterizations provides a detailed picture of the coverage (ca. 1.5 molecules nm$^{-2}$), chemical binding (coexistence of bi- and trifold binding between the phosphonic group and ITO), and arrangement of PA-DAE SAMs on the substrate (tilt angle of 60°). Both XPS and UPS data are further supported by DFT results. Upon alternating illumination of UV and green light, the PA-DAE molecule switches between the open and closed isomeric configuration. This leads to reversible frontier level energy shifts (0.7 eV for HOMO, 0.9 eV for LUMO) as well as to a small work function change (60 meV), evolutions that are fully consistent with the DFT results. Our findings thus provide a pathway to modify the electrode surface electronic properties upon external light illumination, and this strategy will be further employed in ITO-based photoswitchable devices.

**ASSOCIATED CONTENT**

The Supporting Information is available free of charge on the ACS publications website. Unit cell of PA-DAE on ITO, C 1s core level spectra, Quantification of PA-DAE packing density, Quantification of PA-DAE orientation, PA-DAE KPFM image, DFT data of PA-DAE frontier levels (PDF)

**AUTHOR INFORMATION**

**Corresponding Author**

E-mail: nkoch@physik.hu-berlin.de

E-mail: giovanni.ligorio@hu-berlin.de

**Author Contributions**

The manuscript was written through contributions of all authors. All authors have given approval to the final version of the manuscript.

**Funding Sources**

This work was financially supported by the EC through the Marie Curie project ITN iSwitch (GA no. 642196) and the DFG (SFB 951). Computational resources were provided by the Consortium des Équipements de Calcul Intensif (CÉCI) funded by the Belgian National Fund for Scientific Research (F.R.S.-FNRS) under Grant 2.5020.11. J.C. is an FNRS research director.

**Notes**

The authors declare no competing financial interest.